# GPU-based Iterative Cone Beam CT Reconstruction Using Tight Frame Regularization

Xun Jia[1,2], Bin Dong[3], Yifei Lou[4], and Steve B. Jiang[1,2]

[1]Center for Advanced Radiotherapy Technologies, University of California San Diego, La Jolla, CA 92037-0843, USA

[2]Department of Radiation Oncology, University of California San Diego, La Jolla, CA 92037-0843, USA

[3]Department of Mathematics, University of California San Diego, La Jolla, CA 92093-0112, USA

[4]Department of Mathematics, University of California Los Angeles, Los Angeles, CA 90095-1555, USA

E-mail: sbjiang@ucsd.edu

X-ray imaging dose from serial cone-beam CT (CBCT) scans raises a clinical concern in most image guided radiation therapy procedures. It is the goal of this paper to develop a fast GPU-based algorithm to reconstruct high quality CBCT images from undersampled and noisy projection data so as to lower the imaging dose. For this purpose, we have developed an iterative tight frame (TF) based CBCT reconstruction algorithm. A condition that a real CBCT image has a sparse representation under a TF basis is imposed in the iteration process as regularization to the solution. To speed up the computation, a multi-grid method is employed. Our GPU implementation has achieved high computational efficiency and a CBCT image of resolution $512 \times 512 \times 70$ can be reconstructed in ~5 min. We have tested our algorithm on a digital NCAT phantom and a physical Catphan phantom. It is found that our TF-based algorithm is able to reconstrct CBCT in the context of undersampling and low mAs levels. We have also quantitatively analyzed the reconstructed CBCT image quality in terms of modulation-transfer-function and contrast-to-noise ratio under various scanning conditions. The results confirm the high CBCT image quality obtained from our TF algorithm. Moreover, our algorithm has also been validated in a real clinical context using a head-and-neck patient case. Comparisons of the developed TF algorithm and the current state-of-the-art TV algorithm have also been made in various cases studied in terms of reconstructed image quality and computation efficiency.

## 1. Introduction

Cone Beam Computed Tomography (CBCT) is of central importance in cancer radiotherapy. It is particularly convenient for accurate patient setup in image guided radiation therapy (IGRT). Yet, the high imaging dose to healthy organs (a few cGy per scan) (Islam *et al.*, 2006; Kan *et al.*, 2008; Song *et al.*, 2008) in CBCT scans is a clinical concern, especially when CBCT scan is performed before each fraction for the entire treatment course. The imaging dose in CBCT can be reduced by reducing the number of x-ray projections and lowering mAs levels (tube current and pulse duration). In these approaches, however, the consequent CBCT images reconstructed using conventional FDK algorithms (Feldkamp *et al.*, 1984) are highly degraded due to insufficient and noisy projections. It is therefore desirable to develop new techniques to reconstruct high quality CBCT from undersampled and noisy projection data.

Recently, a burst of research in compressed sensing (Donoho and Tanner, 2005; Candes and Romberg, 2006; Candes *et al.*, 2006; Candes and Tao, 2006; Donoho, 2006; Tsaig and Donoho, 2006) have demonstrated the feasibility of recovering signals from incomplete measurements through optimization methods in various mathematical situations. A number of techniques developed in this field have been introduced to the CT or CBCT reconstruction problems from undersampled data (Sidky *et al.*, 2006; Sidky and Pan, 2008; Chen *et al.*, 2008; Cho *et al.*, 2009; Jia *et al.*, 2010b) and have shown their tremendous power in solving such complicated problems. The key idea is that medical images can be sparsely approximated by certain linear transformation and penalizing the $\ell_1$-norm of the image in the transformed domain will enable us to recover the unknown image from highly undersampled data. Using the idea of compressed sensing and sparse approximation of images under transformations to perform CBCT reconstruction has indeed become one of the central topics in medical imaging. Recently, one of the image transformation techniques called tight-frame (TF) transform (Daubechies *et al.*, 2003) has attracted a lot of attentions. These tight frames have the same structure as the traditional wavelets, except that they are redundant systems that generally provide sparser representations to piecewise smooth functions than traditional wavelets. The TF approach is found to be extremely effective and efficient in solving many image restoration problems (Cai *et al.*, 2008; Cai *et al.*, 2009b; Cai *et al.*, 2009a; Cai and Shen, 2010). A short survey on the theory and applications of TF was given by Shen (2010) and a much more detailed survey was given by Dong and Shen (2010). CBCT reconstruction problem can be generally viewed as a 3-dimensional image restoration problem. In such a problem, it has been noted that the discontinuities of the reconstructed piecewise smooth image provide very important information, as they usually account for the boundaries between different objects in the volumetric image. In the TF approach, one tries to restore TF coefficients of the image, which usually correspond to important features, *e.g.* edges, as opposed to the image itself. This allows us to specifically focus on the reconstruction of the important information of the image, hence leading to high quality reconstruction results.

Besides its effectiveness, TF approach also has attractive numerical properties. First, recently invented numerical schemes specifically designed for the TF approach lead to a high convergence rate (Shen *et al.*, 2009; Shen, 2010; Dong and Shen, 2010). Second, the numerical scheme only involves simple matrix-vector or vector operations, making it straightforward to implement the algorithm and parallelize it in a parallel computing structure. It is these numerical properties that lead to high computational efficiency in practice. Moreover, general purpose graphic processing units (GPUs) have offered us a promising prospect of increasing efficiencies of heavy duty tasks in radiotherapy, such as CBCT FDK reconstruction (Xu and Mueller, 2005; Li *et al.*, 2007; Sharp *et al.*, 2007; Xu and Mueller, 2007; Yan *et al.*, 2008), deformable image registration (Sharp *et al.*, 2007; Samant *et al.*, 2008; Gu *et al.*, 2009b), dose calculation (Jacques *et al.*, 2008; Hissoiny *et al.*, 2009; Gu *et al.*, 2009a; Jia *et al.*, 2010a), and treatment plan optimization (Men *et al.*, 2009; Men *et al.*, 2010). Taking advantages of the high computing power of the GPU, the computation efficiency of TF-based CBCT reconstruction is expected to be enhanced considerably.

We have developed a novel CBCT reconstruction algorithm based on TF and implemented it on GPU. The motivation of this work is to provide a new approach for CBCT reconstruction, in addition to the well known FDK-type algorithms and the state-of-the-art iterative reconstruction algorithms, such as total variation (Sidky and Pan, 2008). This work, along with some preliminary validations, will be presented in this paper. Our experiments on a digital phantom, a physical phantom, and a real patient case demonstrate the possibility of reconstructing high quality CBCT images from extremely undersampled and noisy data. The associated high computational efficiency due to the good numerical property of the TF algorithm and our GPU implementation makes this approach practically attractive. Our work, by introducing the novel TF algorithm to the CBCT reconstruction context for the first time, will shed a light to the CBCT reconstruction field and contribute to the realization of low dose CBCT. The rest of this paper is organized as following. Section 2 will describe our method as well as implementation details. In section 3 we will provide the reconstruction results and necessary analysis on the reconstructed volumetric images. Finally, section 4 will conclude our paper.

## 2. Methods

*2.1 Model and Algorithm*

Let us consider a patient volumetric image represented by a function $f(x)$ with $x = (x, y, z) \in \mathbf{R}^3$. A projection operator $P^\theta$ maps $f(x)$ into another function on an x-ray imager plane along a projection angle $\theta$:

$$P^\theta[f](\boldsymbol{u}) = \int_0^{L(\boldsymbol{u})} \mathrm{d}l \, f(\boldsymbol{x}_S + \boldsymbol{n}l) , \qquad (1)$$

where $\boldsymbol{x}_S = (x_S, y_S, z_S)$ is the coordinate of the x-ray source and $\boldsymbol{u} = (u, v) \in \mathbf{R}^2$ is the coordinate of the projection point on the x-ray imager, $\boldsymbol{n} = (n_1, n_2, n_3)$ being a unit vector along the projection direction. Fig. 1 illustrates the geometry. The upper integration limit $L(\boldsymbol{u})$ is the length of the x-ray line. Denote the observed projection image at the angle $\theta$ by $g^\theta(\boldsymbol{u})$. Mathematically speaking, a CBCT reconstruction problem is formulated as to retrieve the volumetric image function $f(\boldsymbol{x})$ based on the observation of $g^\theta(\boldsymbol{u})$ at various angles given the projection mapping in Eq. (1).

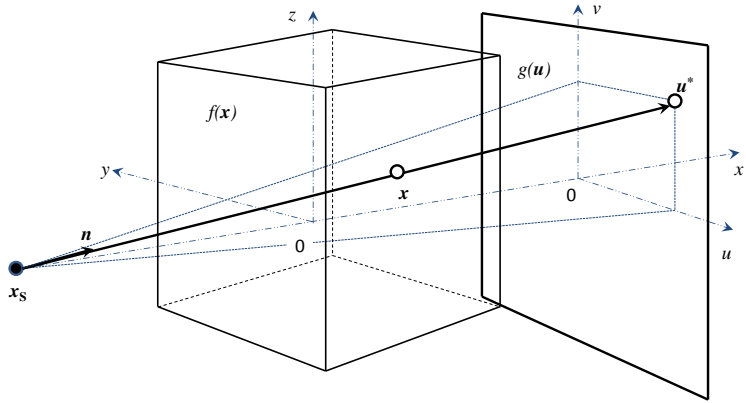

**Figure 1**. The geometry of x-ray projection. The operator $P^\theta$ maps $f(\boldsymbol{x})$ in $\boldsymbol{R}^3$ onto another function $P^\theta[f](\boldsymbol{u})$ in $\boldsymbol{R}^2$, the x-ray imager plane, along a projection angle $\theta$. $L(\boldsymbol{u})$ is the length from $\boldsymbol{x}_S$ to $\boldsymbol{u}^*$ and $l(\boldsymbol{x})$ is that from $\boldsymbol{x}_S$ to $\boldsymbol{x}$. The source to imager distance is $L_0$.

The CBCT image reconstruction from few projections is an underdetermined problem. Because of insufficient measurements made at only a few x-ray projections, there are indeed infinitely many functions $f$ satisfying the condition $P^\theta[f](\boldsymbol{u}) = g^\theta(\boldsymbol{u})$. Therefore, regularization based on some assumptions about the solution $f$ has to be performed during the reconstruction process. These regularization-based CBCT reconstruction approaches usually result in solving challenging minimization problems. The most commonly used approach is an alternative iteration scheme, where, within each iteration step, conditions to be satisfied by the solution is imposed one after another. In our problem, there are three conditions that need to be satisfied by the solution, and three key operations will be performed in each iteration step accordingly. These conditions, as well as the operations ensuring them, will be described in the following.

First, the x-ray projection of the reconstructed volumetric image $f(\boldsymbol{x})$ should match the observation $g^\theta(\boldsymbol{u})$. This condition is commonly achieved by solving a linear system $Pf = g$, where $P$ is the matrix representation of the projection operator $P^\theta$, and $f$ and $g$ are vectors corresponding to the solution $f(\boldsymbol{x})$ and the observation $g^\theta(\boldsymbol{u})$, respectively. Nonetheless, since this is a highly underdetermined problem, any numerical scheme tending to directly solve $Pf = g$ is unstable. Instead, in this work we perform a minimization of an energy $E[f] = \|Pf - g\|_2^2$ by using a conjugate gradient least square (CGLS) algorithm. This algorithm is essentially an iterative algorithm, which generates a new solution $f$ given an initial guess $v$. We formally denote this process as $f \leftarrow \text{CGLS}[v]$, and the details regarding the implementation of the CGLS algorithm will be discussed in section 2.2.2. The CGLS algorithm enables us to efficiently solve this minimization problem, and hence ensures the consistency between the reconstructed volumetric image $f(x)$ and the observation $g^\theta(u)$.

Second, we impose a regularization condition to the solution $f(x)$ that it has a sparse representation under a TF system $X = \{\psi_i(x)\}$. The solution $f(x)$ can be decomposed by $X$ into a set of coefficient as $\alpha_i(x) = \psi_i(x) \otimes f(x)$, where $\otimes$ stands for the convolution of two functions. In this paper, we use the piece-wise linear TF basis (Dong and Shen, 2010; Shen, 2010). Specifically, in 1D, the discrete forms of the basis functions are chosen as $h_0 = \frac{1}{4}[1,2,1]$, $h_1 = \frac{\sqrt{2}}{4}[1,0,-1]$, and $h_2 = \frac{1}{4}[-1,2,-1]$, where $h_0$ is known as a low pass filter and the other two are high pass filters. This is because for a given 1D signal, the convolution with $h_0$ leads to its low frequency component, *i.e.* smoothed skeleton, while convolutions with the other two give high frequency oscillatory parts such as noise signals or edges. The 3D TF basis functions used for the CBCT reconstruction problem can be constructed by the tenser product of the three 1D basis functions, *i.e.* $\psi_i(x,y,z) = h_l(x)h_m(y)h_n(z)$, with integers $l, m, n$ chosen from 0, 1, or 2 and $i = 0,1,\ldots,26$. Among these basis functions, the one with $i = l = m = n = 0$ is the low pass filter, while the other 26 functions are high pass ones. Correspondingly, the coefficient $\alpha_0(x)$ is called the low frequency component and the rest belong to high frequency category. The transform from $f(x)$ to the TF coefficient $\alpha_i(x)$ via convolution is a linear operation. To simplify notation, we can denote this transformation in a matrix notation as $\alpha(x) = Df(x)$. We emphasize here that the introduction of the matrix $D$ is merely for the purpose of simplifying notation. In practice, we still compute this transformation via convolution but not matrix multiplication. Conversely, the function $f(x)$ can be uniquely determined given a set of coefficients $\alpha_i(x)$, $i = 0,1,\ldots,26$, by $f(x) = \sum_i \psi_i(-x) \otimes \alpha_i(x)$, which can be denoted as $f(x) = D^T\alpha(x)$.

It has been observed that many natural images have very sparse representations under the TF system $X$, *i.e.* there are only a small proportion of the elements among the coefficients $\alpha(x)$ that are considerably larger in magnitude than the rest of the elements (Dong and Shen, 2010). It is this property that can be utilized a priori to regularize the reconstructed CBCT image. A common way of imposing this condition into the solution $f$ is to throw away some small TF coefficients among those high frequency components $\alpha_i(x)$ for $i = 1,\ldots,26$, as those components usually come from highly oscillatory signals in the reconstructed CBCT image $f$, such as noise. The deletion of these small coefficients not only sharpens edges but also removes noises. Meanwhile, the low pass components $\alpha_0(x)$ should be left unchanged, since it corresponds to the low frequency signals in $f$, more likely originated from the underlying true image.

One intuitive way of achieving this regularization is to compare each $\alpha_i(x)$, $i = 1,\ldots,26$, with a certain threshold level and set those coefficients below the threshold zero. In practice, it is found that a so called vector shrinkage operation usually leads to better image quality (Cai *et al.*, 2011). In this method, it is the $l_2$-norm of the high frequency component $\|\alpha_h(x)\| = \left[\sum_{i=1}^{26} \alpha_i(x)^2\right]^{1/2}$ that determines whether we keep or

6            X. Jia *et al.*

discard them as opposed to each $\alpha_i(\boldsymbol{x})$ individually. Specifically, the operation we perform on the TF coefficients to regularize the CBCT image is

$$\mathcal{T}_\mu \alpha_i(\boldsymbol{x}) = \begin{cases} \alpha_i(\boldsymbol{x}): & \text{if } i = 0 \\ \alpha_i(\boldsymbol{x}) \max\left[1 - \frac{\mu}{\|\alpha_h(\boldsymbol{x})\|}, 0\right] & : \text{ otherwise} \end{cases}, \qquad (2)$$

where $\mu$ is a predetermined threshold being a tuning parameter for the reconstruction problem. It is understood that such an operation is performed voxel-wise. In particular, if $\|\alpha_h(\boldsymbol{x})\| < \mu$ is found at a certain voxel $\boldsymbol{x}$, $\mathcal{T}_\mu \alpha_i(\boldsymbol{x})$ sets all the high frequency components $\alpha_i(\boldsymbol{x})$ to be zero at this voxel. In summary, to impose regularization on a CBCT image, we first decompose $f$ into the TF space, perform a vector shrinkage operation described as in Eq. (2), and finally reconstruct $f$ based on the new coefficients. This process is symbolically denoted as $f \leftarrow D^T \mathcal{T}_\mu D f$.

Third, since the reconstructed CBCT image $f(\boldsymbol{x})$ physically represents x-ray attenuation coefficient at a spatial point $\boldsymbol{x}$, its positivity has to be ensured during the reconstruction in order to obtain a physically correct solution. For this purpose, we also perform a correction step of the reconstructed image $f(\boldsymbol{x})$ by setting its negative voxel values to be zero. Mathematically, this operation is denoted by $f \leftarrow \mathcal{H} f$, where the operation $\mathcal{H}$ stands for a voxel-wise truncation of the negative values in the CBCT image $f$.

In considering all the components mentioned above, we summarize the reconstruction algorithm as in Algorithm A1:

**Algorithm A1:**

Initialize: $f^{(0)} = 0$.

For $k = 0, 1, \ldots$ do the following steps until convergence

1. Update: $f^{(k+1)} = \text{CGLS}[f^{(k)}]$;
2. Shrinkage: $f^{(k+1)} \leftarrow D^T \mathcal{T}_\mu D f^{(k+1)}$;
3. Correct: $f^{(k+1)} \leftarrow \mathcal{H} f^{(k+1)}$.

Note that there is only one tuning parameter $\mu$ in the algorithm. In practice, its value is carefully tuned so that the best image quality can be obtained. An example of how we choose this parameter is provided in Section 3.2.

We would also like to point out that there is indeed a set of rigorous mathematical theories behind the seemingly heuristic algorithm A1. There is in fact a variation form corresponding to this algorithm, in that there exists an energy functional $E[f(\boldsymbol{x})]$, whose minimizer is the reconstructed CBCT image. The algorithm A1 is one of the algorithms that efficiently solve the minimization problem. Yet, presenting the exact formulation of this variation approach and proof the link to the algorithm A1 is beyond the scope of this paper and relative information can be found in Shen (2010), Dong and Shen (2010), and Cai (2011). With a simple modification, the convergence rate of A1 can be enhanced (Shen *et al.*, 2009; Shen, 2010; Dong and Shen, 2010), leading to Algorithm A2 used in our reconstruction problem:

**Algorithm A2:**

> Initialize: $f^{(0)} = f^{(-1)} = 0$, $t^{(0)} = t^{(-1)} = 1.0$,
> For $k = 0,1,...$ do the following steps until convergence
> 1. Compute: $v^{(k)} \leftarrow f^{(k)} + \frac{t^{(k-1)}-1}{t^{(k)}}[f^{(k)} - f^{(k-1)}]$;
> 2. Update: $f^{(k+1)} = \text{CGLS}[v^{(k)}]$;
> 3. Shrinkage: $f^{(k+1)} \leftarrow D^T \mathcal{T}_\mu D f^{(k+1)}$;
> 4. Correct: $f^{(k+1)} \leftarrow \mathcal{H} f^{(k+1)}$;
> 5. Set: $t^{(k+1)} = \frac{1}{2}[1 + \sqrt{1 + 4t^{(k)^2}}]$.

*2.2 Implementation*

In this paper, the CBCT reconstruction problem is solved with the aforementioned algorithm A2 on an NVIDIA Tesla C1060 card. This GPU card has a total number of 240 processor cores (grouped into 30 multiprocessors with 8 cores each), each with a clock speed of 1.3 GHz. It is also equipped with 4 GB DDR3 memory, shared by all processor cores. Utilizing such a GPU card with tremendous parallel computing ability can considerably elevate the computation efficiency. In this section, we describe some key components of our implementation.

*2.2.1 GPU parallelization*

In fact, a number of computationally intensive tasks involved in algorithm A1 and A2 share a common feature, *i.e.* applying a single operation to different part of data elements. For computation tasks of this type, it is straightforward to accomplish them in a data-parallel fashion, namely having all GPU threads running the same operation, one for a given subset of the data. Such a parallel manner is particularly suitable for the SIMD (single instruction multiple data) structure of a GPU and high computation efficiency can be therefore achieved.

Specifically, the following components in A2 fall into this category: 1) We simply parallelize the voxel-wise vector shrinkage in the Step 3 and the positivity correction of the CBCT image in the Step 4 with one GPU thread responsible for one voxel. 2) The transformation of a CBCT image $f$ into the TF space is merely a convolution operation $\alpha_i(x) = \psi_i(x) \otimes f(x)$. This computation can be performed by having one GPU thread compute the resulted $\alpha_i(x)$ at one $x$ coordinate. The inverse transformation from the TF coefficient $\alpha_i(x)$ to the image $f(x)$ is also a convolution operation and can be achieved in a similar manner. 3) A matrix vector multiplication of the form $g = Pf$ is frequently used in the CGLS method. This operation corresponds to the forward x-ray projection of a volumetric image $f(x)$ to the imager planes, also known as a digital reconstructed radiograph. In our implementation, it is performed in a parallel fashion, with each GPU thread computing the line integral of Eq. (1) along an x-ray line using Siddon's ray-tracing algorithm (Siddon, 1985; Jacobs *et al.*, 1998; Han *et al.*, 1999).

*2.2.2 CGLS method*

Another key component in our implementation is the CGLS solution to the optimization problem $\min_f \|Pf - g\|_2^2$ in Step 2 of A2. In this step, a CGLS method is applied to efficiently find a solution $f^{(k+1)}$ to this least square problem with an initial value of $v^{(k)}$ in an iterative manner (Hestenes and Stiefel, 1952). The details of this CGLS algorithm are given in Appendix 1 in a step-by-step manner. Each iteration step of the CGLS algorithm includes a number of fundamental linear algebra operations. For those simple vector-vector operations and scalar-vector operations, we utilize CUBLAS package (NVIDIA, 2009) for high efficiency. In addition, there are two time-consuming operations requiring special attention, namely matrix-vector multiplications of the form $g = Pf$ or $f = P^T g$, where $P$ is the x-ray projection matrix. Though it is straightforward to accomplish $g = Pf$ on GPU with the Siddon's ray-tracing algorithm as described previously, it is quite cumbersome to carry out a computation of the form $f = P^T g$. It is estimated that the matrix $P$, though being a sparse matrix, contains approximately $4 \times 10^9$ non-zero elements for a typical clinical case studied in this paper, occupying about 16 GB memory space. Such a huge matrix $P$ is too large to be stored in a GPU memory, not to mention computing its transpose. Therefore, a new algorithm for completing the task $f = P^T g$ has to be designed. Without thinking too much, one can compute $f = P^T g$ by still using the Siddon's algorithm. Such an operation, however, is a backward one in that it maps a function $g(\boldsymbol{u})$ on the x-ray imager back to a volumetric image $f(\boldsymbol{x})$ by updating its voxel values along all ray lines. If Siddon's ray-tracing algorithm were still used in the GPU implementation with each thread responsible for updating voxels along a ray line, a memory conflict problem would take place due to the possibility of simultaneously updating a same voxel value by different GPU threads. When this conflict occurs, one thread will have to wait until another thread finishes updating. It is this fact that severely limits the maximal utilization of GPU's massive parallel computing power.

To overcome this difficulty, we analytically compute the explicit form of the resulted volumetric image function $f(\boldsymbol{x})$ when the operator $P^T$ acts on a function $g(\boldsymbol{u})$ on the x-ray imager and obtained a close form expression

$$f(\boldsymbol{x}) = [P^T g](\boldsymbol{x}) = \frac{\Delta x \Delta y \Delta z}{\Delta u \Delta v} \sum_\theta \frac{L^3(\boldsymbol{u}^*)}{L_0 l^2(\boldsymbol{x})} g^\theta(\boldsymbol{u}^*). \tag{3}$$

Here $\boldsymbol{u}^*$ is the coordinate for a point on imager where a ray line connecting the x-ray source at $\boldsymbol{x}_s$ and the point at $\boldsymbol{x}$ intersects with the imager. $L_0$ is the distance from the x-ray source S to the imager, while $l(\boldsymbol{x})$ and $L(\boldsymbol{u}^*)$ are the distance from $\boldsymbol{x}_s$ to $\boldsymbol{x}$ and from $\boldsymbol{x}_s$ to $\boldsymbol{u}^*$ on the imager, respectively. See Fig. 1 for the geometry. $\Delta u$ and $\Delta v$ are the pixel size when we descretize the imager during implementation and $\Delta x$, $\Delta y$, and $\Delta z$ are the size of a voxel. The derivation of Eq. (3) is briefly shown in Appendix 2. Eq. (3), in fact, indicates a very efficient way of performing $f = P^T g$ in a parallel fashion. To compute $f(\boldsymbol{x})$ at a given $\boldsymbol{x}$, we simply take the function values of $g(\boldsymbol{u}^*)$ at the coordinate $\boldsymbol{u}^*$, multiply by proper prefactors, and finally sum over all projection angles $\theta$. In numerical computation, since we always evaluate $g(\boldsymbol{u})$ at a set of discrete coordinates and $\boldsymbol{u}^*$ does not necessarily coincide with these discrete coordinates, a bilinear interpolation is performed to obtain $g^\theta(\boldsymbol{u}^*)$. Now it is ready to perform the parallel computing with each GPU thread for a voxel at a given $\boldsymbol{x}$ coordinate. Extremely high efficiency is expected given the vast parallelization ability of the GPU.

*2.2.3 Multi-grid method*

Another technique we employed to increase computation efficiency is multi-grid method (Brandt, 2002). It has long been known that, the convergence rate of an iterative approach solving an optimization problem is usually worsened when a very fine grid size $\Delta x$, $\Delta y$, and $\Delta z$ is used. Moreover, fine grid also implies a large number of unknown variables, significantly increasing the size of the computation task. A well known multi-grid approach can be utilized to resolve these problems. Suppose we try to reconstruct a volumetric CBCT image $f(\boldsymbol{x})$ on a fine grid $\Omega_h$ of size $h$, we could start with solving the problem on a coarser grid $\Omega_{2h}$ of size $2h$ with the same iterative approach as in Algorithm A2. Upon convergence, we smoothly extend the solution $f_{2h}$ on $\Omega_{2h}$ to the fine grid $\Omega_h$ using, for example, linear interpolation, and use it as the initial guess of the solution on $\Omega_h$. Because of the decent quality of this initial guess, only a few iteration steps of Algorithm A2 are adequate to achieve the final solution on $\Omega_h$. This idea can be further used while seeking the solution $f_{2h}$ by going to an even coarser grid of size $4h$. In practice, we employed a 3-level multi-grid scheme, *i.e.* the reconstruction is sequentially achieved on grids $\Omega_{4h} \to \Omega_{2h} \to \Omega_h$.

*2.3 Comparison with TV reconstruction algorithm*

In the following reconstruction cases, we have also compared our proposed TF-based reconstruction algorithm with the current state-of-the-art iterative CBCT reconstruction algorithm, which uses total variation (TV) as regularization (Sidky and Pan, 2008; Jia *et al.*, 2010b). Specifically, the TV method we employed in this paper tries to reconstruct a CBCT image by solving an optimization problem $f = \mathrm{argmin}_f \frac{1}{2}\|Pf - g\|_2^2 + \mu_{TV}\|f\|_{TV}$, where $\|f\|_{TV} = \int \|\nabla f\|_1 \mathrm{d}x$ is the TV-semi norm. In our implementation, a forward-backward splitting algorithm akin to the Algorithm A1 is used to solve this problem. The main difference from A1 is that the Step 2 becomes solving a sub-problem of $\mathrm{argmin}_f \|f - f^{(k+1)}\|_2^2 + \mu_{TV}\|f\|_{TV}$, also known as a Rudin-Osher-Fatemi (ROF) model (Rudin *et al.*, 1992). Solving this model is achieved by a simple gradient descent method due to the non-existence of a closed form solution. As such, this step becomes an iterative process by itself. We update the solution of this sub-problem along the negative gradient direction in each step with an adaptively adjusted step length. This process terminates when the energy function value decreases less than a certain amount, for instance, 0.1%, in two successive steps. The TV method is also implemented on GPU with the aforementioned multi-grid reconstruction scheme.

Details regarding this algorithm have been previously presented by Jia *et. al.*(Jia *et al.*, 2010b). To ensure the fairness of this comparison, the parameter $\mu_{TV}$ is adjusted manually for each case studied, so that the best image quality can be obtained.

## 3. Experimental Results

In this section, we present the CBCT reconstruction results on a NURBS-based cardiac-torso (NCAT) phantom (Segars *et al.*, 2001), a Catphan phantom (The Phantom Laboratory, Inc., Salem, NY), and a real patient at head-and-heck region. All of the reconstructed CBCT images are of a resolution $512 \times 512 \times 70$ voxels with the voxel size chosen as $0.88 \times 0.88 \times 2.0$ mm$^3$. The x-ray imager resolution is $512 \times 384$ covering an area of $40 \times 30$ cm$^2$. The reconstructed images are much shorter than the imager dimension along the z-direction due to the cone beam divergence. The x-ray source to axes distance is $100$ cm and the source to detector distance is $150$ cm. All of these parameters mimic realistic configurations in a Varian On-Board Imager (OBI) system (Varian Medical Systems, Palo Alto, CA). In all cases we studied, a total number of 40 x-ray projections are used to perform the reconstruction. For the digital NCAT phantom, x-ray projections are numerically computed along 40 equally spaced projection angles covering a full rotation with Siddon's ray tracing algorithm (Siddon, 1985; Han *et al.*, 1999; Jacobs *et al.*, 1998). As for the Catphan phantom case and the real patient case, they are scanned in the Varian OBI system under a full-fan mode in an angular range of $200°$. 363~374 projections are acquired and a subset of 40 equally spaced projections is selected for the reconstruction.

Though reducing radiation dose can be achieved by both reducing mAs and number of projections, in this paper, we focus our work on sparse view (*i.e.* 40 projections) reconstruction for the consideration of computational efficiency. In our algorithm, the length of the measurement *g* and the number of rows of the matrix *P* are linearly proportional to the number of projections. Increasing the projection number will therefore considerably enlarge the problem size and hence prolong the computation time per iteration. Yet, it has been demonstrated that for a given dose level superior image quality with less streaking artifacts is obtained by reducing the radiation dose per projection compared with reducing the number of projections (Tang *et al.*, 2009). So for a given dose but more number of projections, less iteration steps are probably needed to achieve a certain image quality and hence the total computation time may not be prolonged. However, according to our experiments, it is found that reducing the number of projections is more efficient in terms of shortening the computation time. In some clinical applications, such as positioning a patient in radiotherapy, computation efficiency is an important factor to evaluate the feasibility of a reconstruction algorithm. We therefore focus our study on those cases with reduced number of projections in this paper.

*3.1 NCAT phantom and Catphan phantom*

We first test our reconstruction algorithm with a digital NCAT phantom. It is generated at thorax region with a high level of anatomical realism (*e.g.*, detailed bronchial trees). In this simulated case, the projection data are ideal, in that it does not contain contaminations due to noise and scattering as in real scanning. Under this circumstance, a powerful reconstruction algorithm should be able to reconstruct CBCT image almost exactly. For example, Sidky *et. al.* have shown that the TV method can yield accurate reconstruction from very few views (Sidky and Pan, 2008). To test the TF algorithm, we first perform the reconstruction with a large number of iterations (10~30 iterations in one multi-grid level) to get high image quality. The central slice of the reconstructed CBCT image and the ground truth image are presented in Fig. 2 (a) and (c), respectively. Additionally, we also plot profiles along a horizontal and a vertical cut in this slice for the reconstructed image, the ground truth image, as well as the absolute error between them in Fig. 2 (d) and (e). Clearly, the reconstruction error mainly occurs at the boundary of the images, where the intensity changes dramatically. To quantify the reconstruction accuracy in this case, we compute the relative root mean square (RRMS) error as $e = \|f - f_0\|_2 / \|f_0\|_2$, where $f$ is the reconstructed image and $f_0$ is the ground truth one. It is found that the reconstructed 3D volumetric CBCT image attains an RRMS error of $e = 3.27\%$ in this case. If we only compute the RRMS error in the phantom region, *i.e.* excluding those background outside the patient, the RRMS error is $e = 3.06\%$. These numbers clearly demonstrate the ability of the TF algorithm to reconstruct high quality CBCT images in this ideal case. For a comparison purpose, we also present the reconstruction result using the TV method in Fig. 2 (b), (f), and (g), where similar image quality is observed. The RRMS error is $e = 4.14\%$ for the whole image and $e = 4.04\%$ for the region excluding the background.

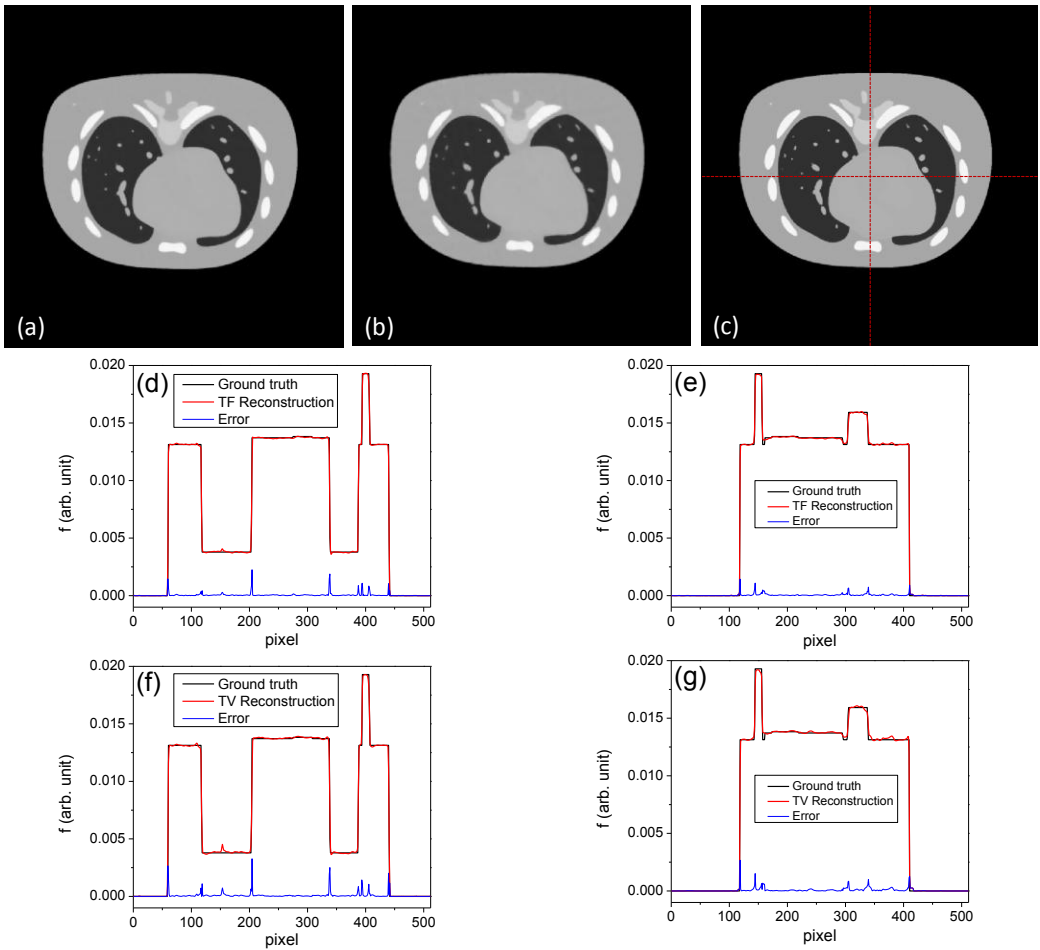

**Figure 2.** The central slice of the reconstructed NCAT phantom by (a) TF method, (b) TV method, and (c) the ground truth image. Dash lines indicate where the profiles in bottom rows are taken. (d) and (e) show the comparisons of the image profiles between the reconstructed image and the ground truth image along a horizontal cut and a vertical cut by TF method, while (f) and (g) are for TV method. The absolute error is also plotted.

It is worth mentioning that the reconstruction time for this case is about 10~20 min on an NVIDIA Tesla C1060 card. In practice, CBCT is mainly used for the patient alignment purpose in cancer radiotherapy, where a fast reconstruction is of essential importance. Though this 10 min reconstruction time has been a big improvement compared with those currently available similar iterative CBCT reconstruction algorithms on CPU, it does not satisfy the requirement in real clinical practice. The above study only serves a purpose of demonstrating the feasibility of using TF as a regularization approach to reconstruct CBCT in an ideal context. In some clinical practice, such as for positioning a patient in cancer radiotherapy, it is adequate to perform less number of iterations for fast image reconstruction, while still yielding acceptable image quality. For this purpose, in the rest of this paper we focus our study to the reconstruction results completed within a given number of iteration steps. In particular, excepted stated otherwise, the iteration steps on the three multi-grid levels are 5, 10, and 15 from the coarsest grid to the finest grid, respectively. This will control the total computation time in about 5~6 min. Same requirements on the number of iterations apply to the state-of-the-art TV-based iterative reconstruction algorithm to make a fair comparison.

Under this condition, the reconstructed CBCT images for the NCAT phantom at the central transverse slice using various algorithms are shown in Fig. 3. We have also scanned a Catphan phantom using Varian OBI at 1.0 mAs/projection and one slice of the resolution phantom is displayed in Fig. 4. First of all, clear streak artifacts are observed in the images produced by the conventional FDK algorithm due to the insufficient number of projections. In contrast, both the TV algorithm and the TF algorithm are able to reconstruct high quality CBCT images even under this extremely under-sampling circumstance and limited number of iteration steps.

While comparing the TV and the TF methods, the image qualities are quite similar, though they are slightly different showing different types of artifacts unique to these two methods. For the TV method, it tends to produce a CBCT image with a high degree of smoothness due to the explicit penalty on the image gradient in the TV term. Meanwhile, the edges and small structures in the image are blurred to a certain extent as a consequence. On the other hand, the TF method is more capable of capturing fine anatomical structures and producing sharper edges. Yet, unlike the TV method, TF penalize the image smoothness in an indirect way, *i.e.* through TF coefficients. This relatively weak control on the image smoothness causes some small but visible residual

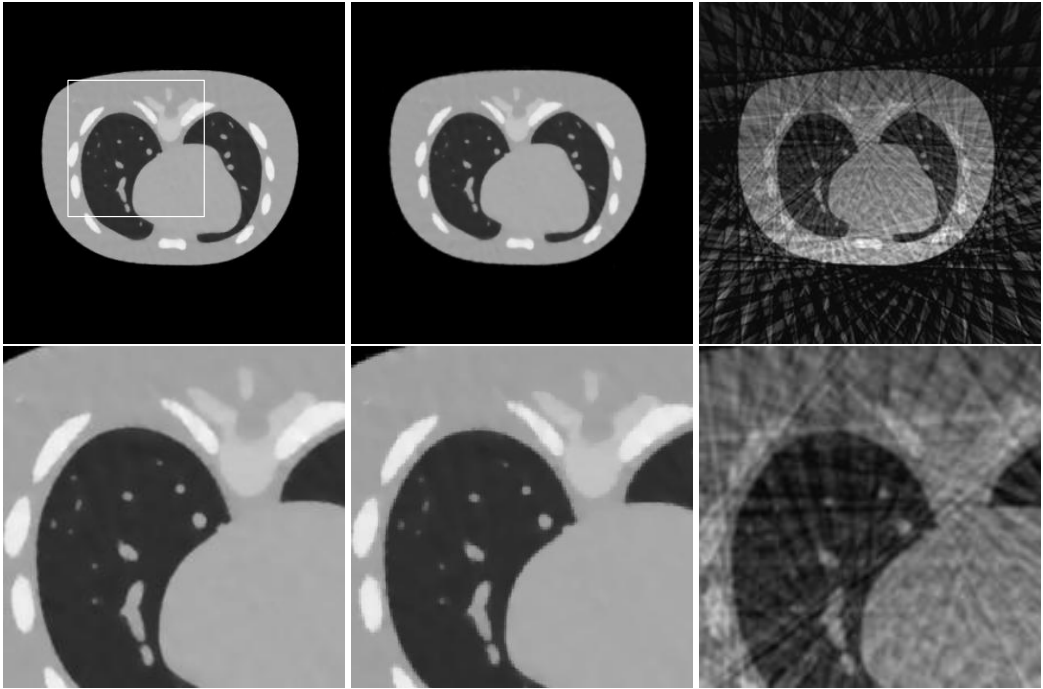

**Figure 3.** One transverse slice of the reconstructed CBCT images for the digital NCAT phantom from 40 projections. Top row, from left to right: images reconstructed using TF algorithm, TV algorithm, and FDK algorithm; Bottom row: zoom in view of the square area for the corresponding images.

streaks, though those streaks are suppressed considerably compared to the FDK results.

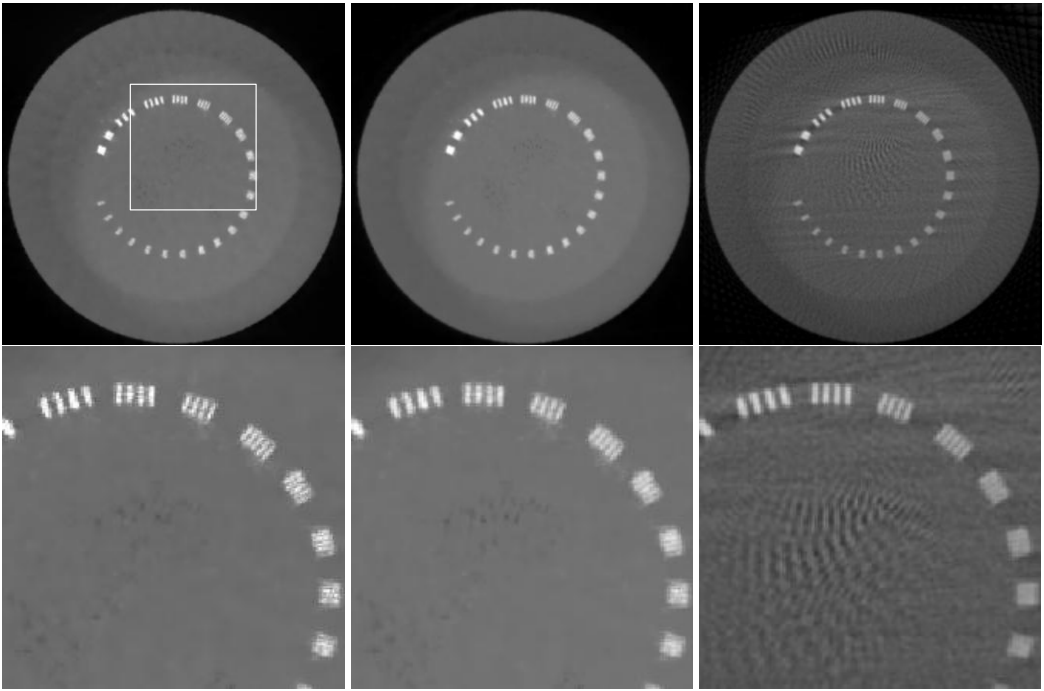

**Figure 4.** One transverse slice of the reconstructed CBCT images for the physical Catphan phantom from 40 projections at 1.0 mAs/projection. Top row, from left to right: images reconstructed using TF algorithm, TV algorithm, and FDK algorithm; Bottom row: zoom in view of the square area for the corresponding images.

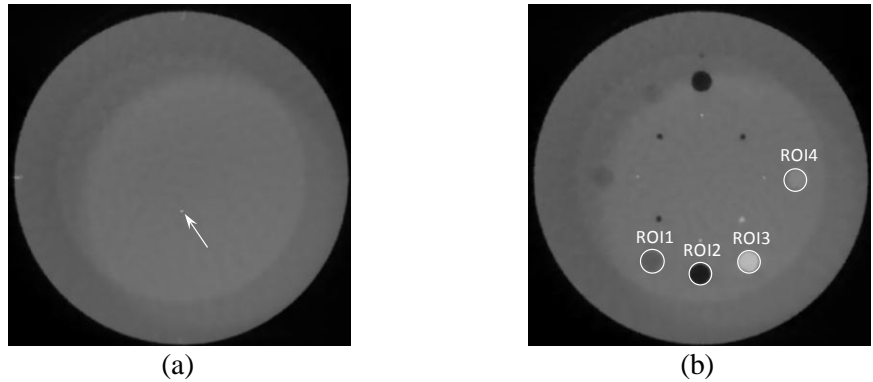

**Figure 5.** (a) A transverse slice of the Catphan phantom used to measure MTF. (b) A transverse slice of the Catphan phantom used to measure $CNR$.

For instance, there are still some residual streak artifacts seen on the Catphan phantom case reconstructed by the TF method, but the image sharpness is improved at the lung edges and the small structures inside the lung are less blurred in the NCAT phantom. It is also worth noticing that these artifacts found in the TF and the TV results are mainly due to the early termination of the reconstruction process. If it is allowed to perform the iteration for a much longer them, both methods can remove their corresponding artifacts to a fairly good extent, as indicated in Fig. 2 for the TF method and in previous studies by Sidky *et. al.* (Sidky and Pan, 2008) for the TV method.

*3.2 Quantitative analysis*

The Catphan phantom contains a layer consisting of a single point-like structure of a diameter $0.28 \, \mathrm{mm}$, see Fig. 5(a). This structure enables us to measure the in plane modulation transfer function (MTF) of the reconstructed CBCT images, which characterizes the spatial resolution inside the transverse plane. For this purpose, we crop a square region of size $21 \times 21 \, \mathrm{pixel}^2$ in this slice centering at this structure. After subtracting the background, we compute the point spread function. The MTF is obtained by first performing 2D fast Fourier transform and then averaging the amplitude along the angular direction.

First, at a constant mAs level of $1.0 \, \mathrm{mAs/projection}$, we compare the spatial resolution in the images reconstructed by the TF, the TV, and the FDK algorithms. Fig. 6(a) presents the patch images containing the dotted structure and the corresponding measured MTF curves. Apparently, the structure is blurred most by the FDK algorithm, and slightly more by the TV method than by the TF method. As a consequence, the TF method results in the best MTF curve among all three methods and therefore yields the highest spatial resolution on the reconstructed images. Second, for the TF method, we compare the resolution at different mAs levels and the results are depicted in Fig. 6(b). As expected, the spatial resolution is deteriorated when low mAs level scan is used due to more and more noise component induced in the x-ray projections. Especially, at an extremely low mAs level of 0.1 mAs/projection, the dotted structure is almost not resolved. For comparison, the TV results are shown in Fig. 6(c). Again, the resolution degrades as the mAs level is reduced. At the low mAs level of 0.10mAs/projection, it is also found that the spatial resolution of TV results is slightly higher than that of the TF method.

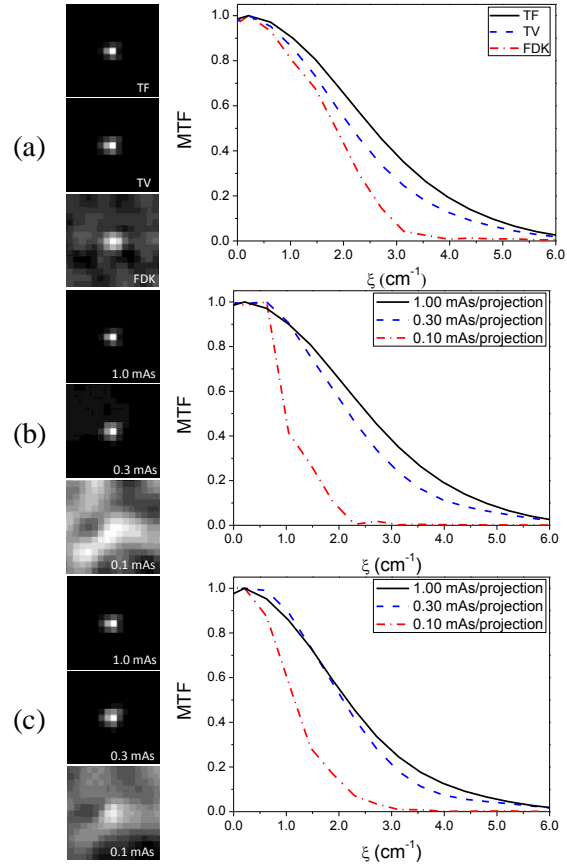

**Figure 6.** (a) Three patches used to measure MTF and the corresponding MTF curves in CBCT images reconstructed from TF, TV, and FDK algorithms at 1.0 mAs/projection with 40 projections. (b) and (c) Three patches used to measure MTF and the corresponding MTF curves in CBCT images reconstructed from TF method and TV method at 1.0, 0.3, and 0.1 mAs/projections with 40 projections.

In order to quantitatively evaluate the contrast of the reconstructed CBCT images, we measure contrast-to-noise ratio ($CNR$). For a given region of interest (ROI), $CNR$ is calculated as $CNR = 2|S - S_b|/(\sigma + \sigma_b)$, where $S$ and $S_b$ are the mean pixel values over the ROI and in the background, respectively, and $\sigma$ and $\sigma_b$ are the standard deviation of the pixel values inside the ROI and in the background. Before computing the $CNR$, a key observation is that $CNR$ is affected by the parameter $\mu$ which controls to what extent we would like to regularize the solution via the TF term. In fact, a small amount $\mu$ is not sufficient to regularize the solution, leading to a high noise level and hence a low $CNR$. In contrast, a large $\mu$ tends to over-smooth the CBCT image and reduce the contrast between different structures. Therefore, there exists an optimal $\mu$ level in the reconstruction. Take the case at 1.0 mAs/projection and 40 projections as an example, we perform CBCT reconstruction with different $\mu$ values and compute the $CNR$s for the four

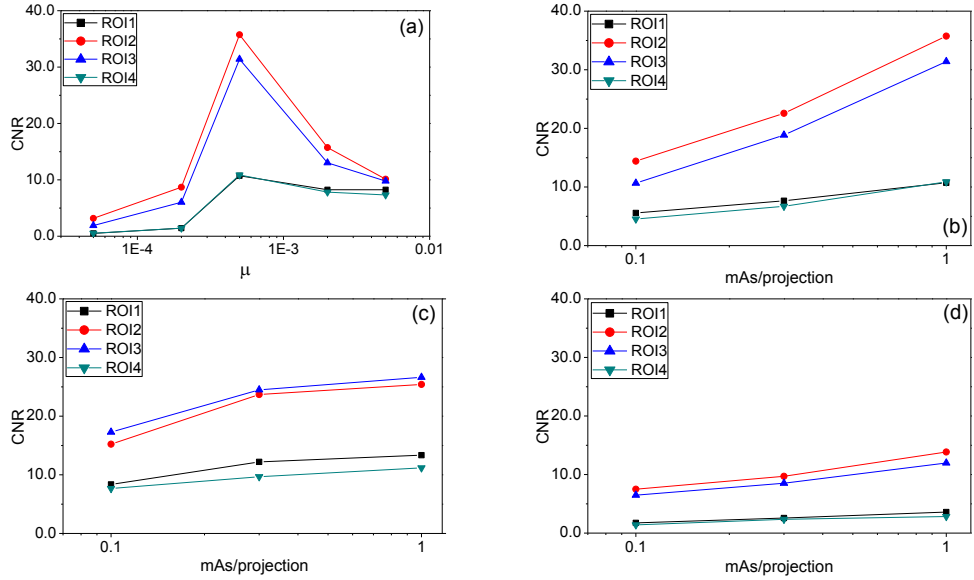

**Figure 7.** (a) $CNR$s at various ROIs as functions of the parameter $\mu$ at 1.0 mAs/projection and 40 projections. (b)~(d) $CNR$s computed at various ROIs as functions of mAs levels at 40 projections reconstructed using our TF algorithm, the TV algorithm and the FDK algorithm.

ROIs indicated in Fig.5(b). The results are shown in Fig. 7(a). Clearly, the best $CNR$s are achieved for $\mu \sim 5.0 \times 10^{-4}$. In principle, the optimal parameter would depend on the noise level in the input projection data, which is a function of the system parameters such as mAs levels, number of projections, reconstruction resolution *etc.* as well as the object being scanned. The precise establishment of the relationship between the optimal $\mu$ value and each of the aforementioned factors will be studied in our future work. Throughout this paper, all the reconstruction cases are performed under the optimal $\mu$ values except stated explicitly.

In Fig. 7(b)~(d), we plot the dependence of $CNR$s on mAs levels measured in those four ROIs in the CBCT images reconstructed using various algorithms. As expected, a higher $CNR$ can be achieved when a higher mAs level is used in the CBCT scan, and hence all of the curves generally follow a monotonically increasing trend. FDK algorithm attains the lowest $CNR$ levels due to the absence of image regularization. As for the TF algorithms, though relatively high $CNR$s can be achieved in high mAs cases, the $CNR$s decrease with mAs sharply. In contrast, the TV algorithm maintains the $CNR$ levels better than the TF algorithm and attains higher $CNR$s at low mA cases, indicating its superior ability of controlling noise at low mAs contexts.

*3.3 Patient case*

Finally, we present our TF-based CBCT reconstruction results on realistic head-and-neck anatomical geometry. A patient's head-and-neck CBCT scan is taken using a Varian OBI system with 0.4 mAs/projection. The reconstruction results using the three reconstruction algorithms with 40 x-ray projections are shown in Fig. 8. Due to the complicated geometry and high contrast between bony structures, dental filling, and soft tissues in this head-and-neck region, streak artifacts are severe in the images obtained from FDK algorithm. On the other hand, the TV algorithm and the TF algorithm both can capture the main anatomical features and suppress the streaking artifacts, while the boundaries, especially of those bony structures, are blurred to a certain extent. It is found the TV algorithm can suppress the streaks better by comparing the residual streaks around the dental filling. One the other hand, the TF method leads to visually slightly sharper boundaries of the bony structures.

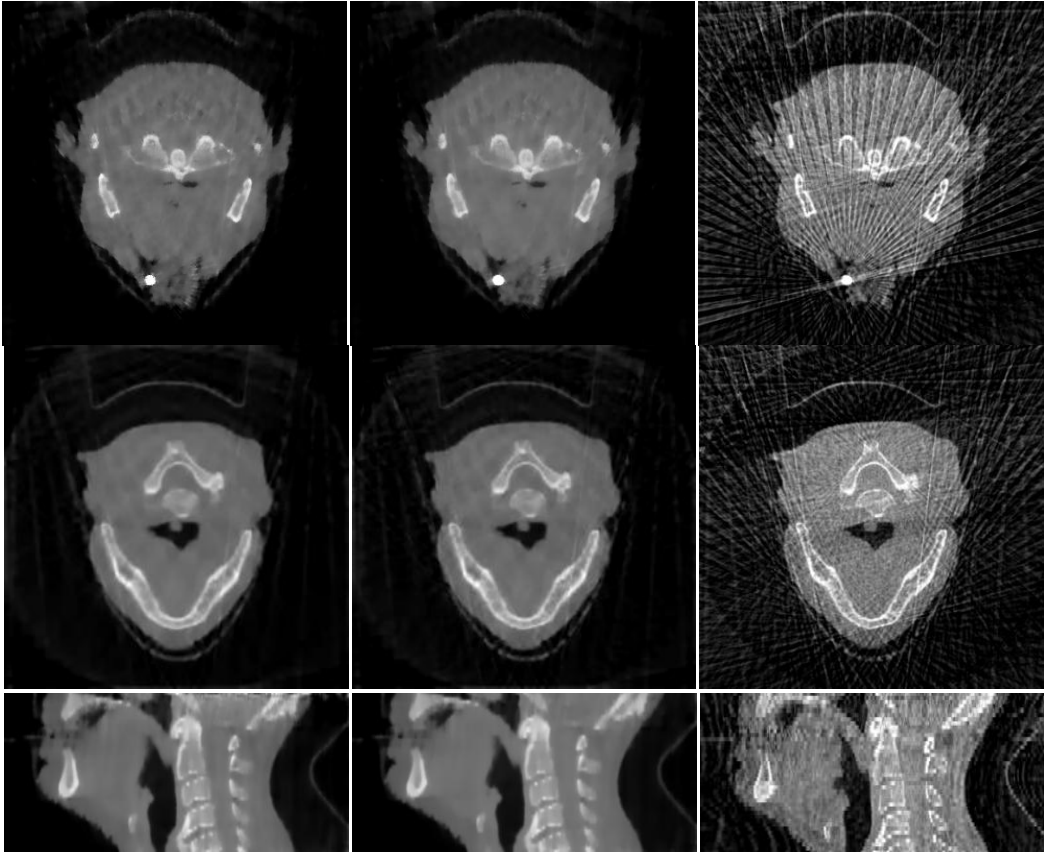

**Figure 8.** Two transverse slices and one sagittal slice of a real head-and-neck patient CBCT reconstructed from the TF algorithm (first column), the TV algorithm (second column), and the FDK algorithm (third column) using 40 projections.

**4. Conclusion and Discussions**

In this paper, we have developed a TF-based fast iterative algorithm for CBCT reconstruction. By iteratively applying three steps to impose three key conditions that a reconstructed CBCT image should satisfy, we can reconstruct CBCT images with undersampled and noisy projection data. In particular, the underline assumption that a real CBCT image has a sparse representation under a TF basis is found to be valid and robust in the reconstruction, leading to high quality results. In practice, due to the GPU implementation, the multi-grid method, and various techniques we employed, high compuational efficiecny has been achieved. We have tested our algorithm on a digital NCAT phantom, a physical Catphan phantom. Quantitative analysis of the CBCT image quality has been performed with respect to the MTF and $CNR$ under various scanning cases, and the results confirm the high CBCT image quality obtained from our TF algorithm. Moreover, reconstructions performed on a head-and-neck patient have presented very promising results in real clinical applications. In our future work, we plan to perform systematical studies to assess the clinical gain of this new algorithm over existing algorithms using a large set of representative patient images and a set of clinically relevant metrics.

It is not quite surprising that our TF based iterative reconstruction algorithm outperforms the FDK algorithm in the undersampling context. But it is of importance and interest to compare the TF algorithm with the current state-of-the-art iterative CBCT reconstruction algorithm, namely TV. In addition to the comparison results shown in the paper, we provide some further discussions regarding these two methods here. However, since the following points are made based on our initial studies on only a few cases presented in this paper, they are by no means conclusive. Further investigation regarding the systematical comparison between the two methods is certainly one of our central topics in near future.

First, in all the cases studied in this paper, the image quality from TF and TV algorithms are quite similar. Yet, there are some visible differences between those results showing the unique characteristics of those two algorithms. Since the TV method explicit penalizes the image gradient via the TV term, it tends to produce a CBCT image with a high degree of smoothness. Meanwhile, the edges in an image are usually blurred to a certain extent as a consequence. In contrast, the TF method regularizes an image and enforces smoothness in an indirect manner, *i.e.* through TF coefficients. It is therefore capable of preserving sharper edges to a better degree, though some residual artifacts are often seen in the reconstructed images. One should keep in mind that these artifacts unique to those two algorithms occur in the intermediate stage of the iteration, as we only performed a few number of iterations to reconstruct those testing cases for the consideration of controlling computation time. If reconstructions with a large number of iterations are allowed, both TV and TF are capable of removing their own artifacts to a satisfactory degree, as having been demonstrated in Fig. 2 for the TF algorithm and in many other studies for the TV algorithm. Under that circumstance, the difference between the reconstruction results produced by the two algorithms is expected to be diminishing.

Second, TV and TF methods show different efficacy in terms of balancing contrast and noise, which result in different characteristics in the $CNR$ plot. Specifically, it is found in Fig. 7 that TF leads to higher $CNR$s in high mAs cases, while TV achieves higher $CNR$s in the low mAs limit. It is naturally expect that TV can result in a very high $CNR$, as it suppresses noise very well. This is, however, not quite the case sometimes due to the loss of contrast. Since TV solves a minimization sub-problem of $\mathrm{argmin}_f \|f - f^{(k+1)}\|_2^2 + \mu_{TV} \|f\|_{TV}$, *i.e.* the ROF model, in each iteration, a certain amount of contrast is usually lost in this process. This fact has been observed in many studies and even mathematically demonstrated (Meyer, 2001). In contrast, TF method panelizes only high frequency components, while leave low frequency components unchanged. The unaltered low frequency components serve as a skeleton of the reconstructed image, which maintains the image contrast. It is this fact that TV achieves relatively lower $CNR$s in the high mAs cases. On the low mAs limit, where a large amount of noise signal appears in the projections, TV start to demonstrate its superior ability of controlling noise in the reconstructed images relative to the TF algorithm, leading to higher $CNR$s despite the loss of contrast.

Third, the computation efficiency is found different. In our reconstruction, it is observed that the absolute computation time per iteration step is 1.1, 4.1, and 15.2 sec for the TF algorithm on the three multi-scale levels and the corresponding time for TV is about 1.8, 5.1, and 17.7 sec. Among each iteration, CGLS update accounts for about 70% of the computation time. Comparing TV and TF algorithms, the main difference is at the stage of performing image regularization. TV solves an ROF model in this sub-problem, while TF uses a deterministic way of thresholding the TF coefficients. Since the ROF model is solved with a simple gradient descent method, which is an iterative algorithm by itself, the performance is relatively lower than the deterministic way employed in the TF algorithm. Yet, there exist some novel algorithms that solve the ROF model very fast given the current development on image processing. For instance, the split Bregman algorithm (Goldstein and Osher, 2009) has been shown to be capable of solving the ROF model with an high efficiency, though we did not exploit the possibility of integrating this algorithm in this work.

Last but not least, we would like to point an interesting connection between the TV algorithm and the TF algorithm. When computing the TF coefficient by convolving a signal $f$ with a high pass filter, such as $h_1 = \frac{\sqrt{2}}{4}[1, 0, -1]$, the result is essentially an approximation of the partial derivative with a central finite difference scheme. Therefore, utilizing the information of $\|\alpha_h(x)\| = \left[\sum_{i=1}^{26} \alpha_i(x)^2\right]^{1/2}$ to enforce the image smoothness is to some degree similar to using a TV term of $\sqrt{\left(\frac{\partial f}{\partial x}\right)^2 + \left(\frac{\partial f}{\partial y}\right)^2 + \left(\frac{\partial f}{\partial z}\right)^2}$ in its discrete format. Though this argument is not rigorous, the connection between the TV algorithm and the TF algorithm has been recently mathematically established under a certain conditions (Cai *et al.*, 2011).

## Acknowledgements

This work is supported in part by the University of California Lab Fees Research Program and by Varian Medical Systems. We would like to thank NVIDIA for providing GPU cards for this project.

20                                                                      X. Jia *et al.***Appendix**

**1. CGLS algorithm**

CGLS algorithm (Hestenes and Stiefel, 1952) solves the least-square problem $\min_x \|Px - y\|_2^2$ in an iterative manner using conjugate gradient method. Specifically, the algorithm performs following iterations:

**Algorithm CGLS:**

Initialize: $x^{(0)}$; $r^{(0)} = y - Px^{(0)}$; $p^{(0)} = s^{(0)} = P^T r^{(0)}$; $\gamma^{(0)} = \|s^{(0)}\|_2^2$;

For $l = 0,1,\ldots,M$, do the following steps
1. $q^{(l)} = Pp^{(l)}$
2. $\alpha^{(l)} = \gamma^{(l)}/\|q^{(l)}\|_2^2$;
3. $x^{(l+1)} = x^{(l)} + \alpha^{(l)}p^{(l)}$, $r^{(l+1)} = r^{(l)} - \alpha^{(l)}q^{(l)}$;
4. $s^{(l+1)} = P^T r^{(l+1)}$;
5. $\gamma^{(l+1)} = \|s^{(l+1)}\|_2^2$;
6. $\beta^{(l)} = \gamma^{(l+1)}/\gamma^{(l)}$;
7. $p^{(l+1)} = s^{(l+1)} + \beta^{(l)}p^{(l)}$.

Output $x^{(M+1)}$ as the solution.

Noticing that in the context of CBCT reconstruction with only a few projections, the normal equation $P^T Px = P^T y$ is indeed underdetermined. The convergence of the CGLS algorithm for underdetermined problems have been studied previously (Kammerer and Nashed, 1972). In our reconstruction algorithm, the CGLS is used as a means to ensure the data fidelity condition during each iteration step of the reconstruction. Specifically, given an input image $x^{(0)} = f$, the CGLS algorithm outputs a solution $f' = x^{(M+1)}$ which is better than the input in the sense that the residual $\|Pf' - y\|_2^2$ is smaller than $\|Pf - y\|_2^2$. This fact always holds regardless the singularity of the linear system.

Since the use of CGLS is merely for ensuring data fidelity via minimizing the residual $l_2$ norm, in each outer iteration of our TF algorithm, it is not necessary to perform CGLS iteration till converge. In practice, $M = 2 \sim 3$ CGLS steps in each outer iteration step is found sufficient. This approach is also favorable in considering the computation efficiency, as more CGLS steps per outer iteration step will considerably slow down the overall efficiency.

**2. Derivation of Eq. (3)**

Let $f(.): \mathbf{R}^3 \to \mathbf{R}$ and $g(.): \mathbf{R}^2 \to \mathbf{R}$ be two smooth enough functions in the CBCT image domain and in the x-ray projection image domain, respectively. The operator $P^{\theta^T}$, being the adjoint operator of the x-ray projection operator $P^\theta$, should satisfy the condition

$$\langle f, P^{\theta^T} g \rangle = \langle P^\theta f, g \rangle, \tag{A1}$$

where $\langle .,. \rangle$ denotes the inner product. This condition can be explicitly expressed as

$$\int \mathrm{d}\pmb{x}\, f(\pmb{x}) P^{\theta^T}[g](\pmb{x}) = \int \mathrm{d}\pmb{u}\, P^{\theta}[f](\pmb{u}) g(\pmb{u}). \tag{A2}$$

Now take the functional variation with respect to $f(\pmb{x})$ on both sides of Eq. (A2) and interchange the order of integral and variation on the right hand side. This yields

$$P^{\theta^T}[g](\pmb{x}) = \frac{\delta}{\delta f(\pmb{x})} \int \mathrm{d}\pmb{u}\, P^{\theta}[f](\pmb{u}) g(\pmb{u}) = \int \mathrm{d}\pmb{u}\, g(\pmb{u}) \frac{\delta}{\delta f(\pmb{x})} P^{\theta}[f](\pmb{u}). \tag{A3}$$

With help of a delta function we could rewrite Eq. (1) as

$$P^{\theta}[f](\pmb{u}) = \int \mathrm{d}l \mathrm{d}\pmb{x}\, f(\pmb{x})\, \delta(\pmb{x} - \pmb{x}_S - \pmb{n}l). \tag{A4}$$

Now substituting (A4) into (A3), we obtain

$$P^{\theta^T}[g](\pmb{x}) = \int \mathrm{d}l \mathrm{d}\pmb{u}\, g(\pmb{u})\, \delta(\pmb{x} - \pmb{x}_S - \pmb{n}l) = \frac{L^3(\pmb{u}^*)}{L_0 l^2(\pmb{x})} g(\pmb{u}^*), \tag{A5}$$

where $\pmb{u}^*$ is the coordinate of a point on imager, at which a ray line connecting the source $\pmb{x}_S$ and the point $\pmb{x}$ intersects with the imager. $L(\pmb{u}^*)$ is the length from $\pmb{x}_S$ to $\pmb{u}^*$ and $l(\pmb{x})$ is that from $\pmb{x}_S$ to $\pmb{x}$. The source to imager distance is $L_0$. Additionally, a summation over projection angles $\theta$ is performed in Eq. (3) to account for all the x-ray projection images.

One caveat when implementing (A5) is that this equation is derived from condition (A1), where the inner product of two functions is defined in an integral sense. In the CGLS algorithm, both $P$ and $P^T$ are viewed as matrices. Therefore an inner product defined in the vector sense, i.e. $\langle f, g \rangle = \sum_i f_i g_i$ for two vectors $f$ and $g$, should be understood in (A1). Changing the inner product from a function form to a vector form results in a numerical factor in Eq. (3), simply being the ratio of pixel size $\Delta u \Delta v$ to the voxel size $\Delta x \Delta y \Delta z$. We have tested the accuracy of such defined operator $P^T$ in terms of satisfying condition expressed in Eq. (A1). Numerical experiments indicate that this condition is satisfied with numerical error less than 1%. Though this could lead to CT number inaccuracy in the reconstructed CBCT image, absolution accuracy of CT number is not crucial in the use of CBCT in cancer radiotherapy, since CBCT is mainly used for patient setup purpose. Meanwhile, the readers should be aware of this potential inaccuracy of the Hounsfield Unit when utilizing Eq. (3).